# The performance limits of epigraphene Hall sensors

H. He[1], N. Shetty[1], T. Bauch[1], S. Kubatkin[1], T. Kaufmann[2], M. Cornils[2], R. Yakimova[3], and S. Lara-Avila[1,4]

[1] *Department of Microtechnology and Nanoscience, Chalmers University of Technology, 412 96 Gothenburg, Sweden.*

[2] *TDK-Micronas GmbH, Hans-Bunte-Strasse 19, D-79108 Freiburg, Germany*

[3] *Department of Physics, Chemistry and Biology, Linkoping University, 581 83 Linköping, Sweden.*

[4] *National Physical Laboratory, Hampton Road, Teddington TW11 0LW, UK*

Epitaxial graphene on silicon carbide, or epigraphene, provides an excellent platform for Hall sensing devices in terms of both high electrical quality and scalability. However, the challenge in controlling its carrier density has thus far prevented systematic studies of epigraphene Hall sensor performance. In this work we investigate epigraphene Hall sensors where epigraphene is doped across the Dirac point using molecular doping. Depending on the carrier density, molecular-doped epigraphene Hall sensors reach room temperature sensitivities $S_V$=0.23 V/VT, $S_I$=1440 V/AT and magnetic field detection limits down to $B_{MIN}$=27 nT/√Hz at 20 kHz. Thermally stabilized devices demonstrate operation up to 150 ˚C with $S_V$=0.12 V/VT, $S_I$=300 V/AT and $B_{MIN}$~100 nT/√Hz at 20 kHz.

Based on the classical Hall Effect, solid-state Hall sensors represent a large portion of magnetometers which are extensively used in automotive, marine and consumer electronics applications. Hall sensors based on silicon see widespread use owing to well-established and low-cost production methods,[1–3] but increasing requirements placed on improved magnetic performance or resilience to harsh conditions like high temperatures, demand the exploration of other, even more suitable materials.[4]

Hall sensors detect magnetic fields by measuring the Hall voltage $V_H$ induced by an external field B. High device sensitivity implies a large magnitude of $V_H$ response to an external field, for a given bias current $I_B$ or voltage $V_B$. This leads to two important material-related metrics: the current-related sensitivity $S_I$=|$V_H$/(B$I_B$)| (V/(AT)), which is essentially determined by the Hall coefficient $R_H$ (Ω/T), and the voltage-related sensitivity $S_V$=|$V_H$/(B$V_B$)| (V/(VT)) which is ultimately limited by the carrier mobility $\mu$ (m$^2$/(Vs)), where $\rho_{XX}$ is sheet resistance.

Graphene appears to be a natural candidate for highly sensitive Hall elements due to its high mobility, and the possibility to tune carrier density *n* down to zero towards charge neutrality (Dirac point). Low carrier density is desirable because it increases the Hall coefficient, $R_H = 1/(ne)$.[5,6] Moreover, since the mobility $\mu = R_H/\rho_{XX}$ of graphene is inversely proportional to carrier density as $\mu \propto 1/\sqrt{n}$,[7] decreasing *n* towards neutrality would increase both $S_I$ and $S_V$. In principle, low *n* leads to an increase in $\rho_{XX}$, which follows the relation $\rho_{XX} \propto 1/n$, in the limit where charged impurity scattering dominates (Supplementary S1).[8,9] Yet, decreasing *n* can actually lead to a lower magnetic field detection limit, $B_{MIN} = V_N/(I_B R_H)$ (T/√Hz), where $V_N$ is the voltage noise spectral density (V/√Hz). If Johnson-Nyquist noise dominates, then $V_N = V_{TH} \propto \sqrt{4k_B T \rho_{XX}}$, with Boltzmann constant $k_B$, temperature T, and the detection limit scales as $B_{MIN} \propto V_N/R_H \propto \sqrt{n}$ for a fixed $I_B$. Disorder in real graphene samples prevents it from reaching true charge neutrality, but high-quality graphene can approach low carrier densities.[10]

The highest quality graphene is obtained by mechanical exfoliation of graphite and encapsulation in hexagonal boron nitride (hBN-G). As Hall sensor, hBN-G has shown ultra-high device sensitivities, and detection limits comparable to that of silicon.[11] However, this approach serves only as a proof of principle of the capabilities of graphene Hall sensors since device fabrication cannot be scaled-up. Graphene grown using chemical vapor deposition (CVD) is a more scalable technology which also can reach high sensitivities, but reported performance varies greatly,[12–14] perhaps due to variability in material growth and the need for subsequent transfer to suitable substrates.[15]

Epitaxial graphene on SiC substrate (epigraphene) is another attractive scalable technology. The insulating substrate allows for direct mass-fabrication of devices over wafer-scales,[16,17] forgoing the need for graphene-transfer thus increasing reproducibility and yield. Epigraphene is also compatible with operation at temperatures exceeding common industrial requirements.[18,19] Despite these advantages, epigraphene remains relatively unexplored for Hall sensing in literature,[18] possibly owing to the difficulties in tuning carrier density due to high intrinsic n-doping, pinned by the substrate.[20–22]

We report the exploration of the performance limits of epigraphene Hall sensors for varying doping levels across the Dirac point. Carrier density control is enabled by a molecular doping method using electron acceptors F4TCNQ assembled on the surface of epigraphene.[23] Devices doped using this method have already shown excellent electrical properties and low charge-disorder, albeit at low temperatures.[24,25] We investigate Hall sensor figures of merit $B_{MIN}$, $S_V$, and $S_I$, and finally thermal stability in ambient conditions from room temperature and just above 200 °C. Furthermore, we establish the limits for optimal operation of epigraphene Hall devices under realistic operational conditions.

Epigraphene was grown on 4H-SiC chips encased in a graphite crucible and heated using RF heating to around 1850 °C in an inert atmosphere of 1 bar argon.[16] Transmission mode microscopy was used to select only samples with over 90 % monolayer coverage.[26] Device fabrication used standard electron beam lithography. Epigraphene is removed using oxygen plasma etching and the metal contacts are deposited using physical vapor deposition of 5 nm Ti and 80 nm Au. The finished device is spin-coated with molecular dopants and the final carrier density is tuned by annealing at T=160 °C, with varying annealing time depending on the desired final doping level.[23] Electrical characterization was performed primarily using the Van der Pauw (VdP) method, with samples measured at room temperature and under ambient conditions unless otherwise stated. A magnetic field perpendicular to the chip surface was applied using a coil electromagnet up to 100 mT). Noise measurements where performed by taking the power spectral density (PSD) using a voltage amplifier DLPVA-100-F-D from Femto, with bandwidth limited to 100 kHz and measured input noise level of 9 nV/√Hz. High-field measurements were performed in PPMS (Quantum design) cryostat (2-300 K) with a superconducting magnet providing fields up to 14 T. For heating experiments, the sample was mounted using epoxy on a ceramic heater, and temperature was monitored using a Pt100-resistor.

Seven epigraphene Hall sensors (Fig. 1(a)), spread across four chips, were investigated in total. They were designed using symmetric square or cross shaped geometries optimized with respect to $S_V$.[27,28] Cryogenic measurements on a molecular-doped sensor demonstrates a full transition to half-integer Quantum Hall regime, with vanishing sheet resistance $\rho_{XX}$ and quantized Hall resistance $R_{XY} = h/2e^2$ (Fig. 1(b)). These measurements verify that the devices are made of high-quality monolayer graphene with uniform doping.

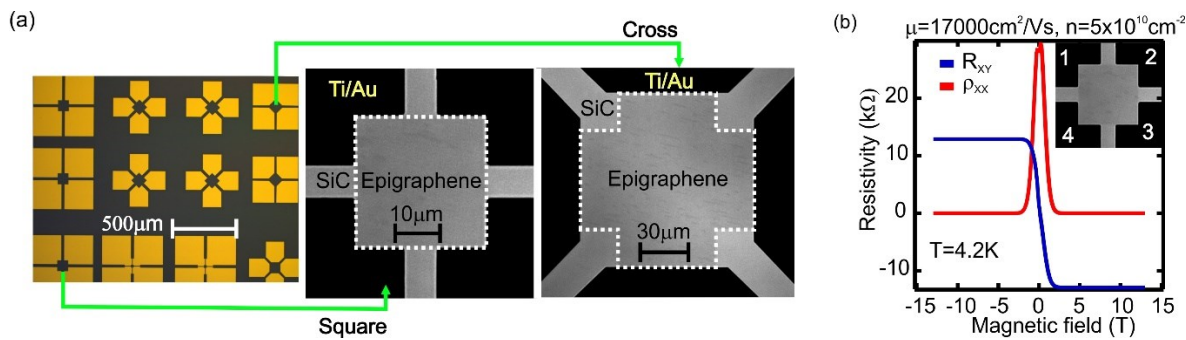

FIG. 1. (a) Optical micrographs of the layout epigraphene Hall sensors. Each chip contains an array of sensors with square and cross shaped geometries. (b) Molecular-doped Hall sensor displays the half-integer quantum Hall effect at cryogenic temperatures. $R_{XY}$ used e.g. contacts 1-3 for bias current and 2-4 to measure Hall voltage. $\rho_{XX}$ used e.g. 1-2 for bias and 4-3 for voltage measurement.

Hall measurements of the transversal resistance $R_{XY} = V_H/I_B$ serve as basis for the evaluation of epigraphene Hall magnetometers. Hall coefficient, carrier densities, and mobilities are calculated from measurements in low magnetic fields (B<0.5 T) as $R_H = dR_{XY}/dB$, $n = 1/(eR_H)$, and $\mu = R_H/\rho_{XX}$, respectively. For the low-field range, the linearity error of $R_{XY}$ is below 1 %, which is determined by the percentage deviation of the raw data from the low-field linear fit (Fig. 2(a)). The samples were tested up to B=13 T at room temperature. For low doping ($R_H$=1284 $\Omega$/T, n=4.9x10$^{11}$ cm$^{-2}$) the transversal resistance remains within 5% error in a range of B=±1.2 T, but for higher doping ($R_H$=949 $\Omega$/T, n=6.6x10$^{11}$ cm$^{-2}$) the 5% error range increases to B=±6 T. Figure 2(b) shows a summary of the carrier densities achieved in our experiments. The gap in data near charge neutrality (n=0) indicates the disordered charge-puddle regime, characterized by a nonlinear low-field $R_{XY}$.[23] At room temperature the maximum measured values of $R_H$ and $\mu$ are $R_H$=1440 $\Omega$/T and $\mu$=2300 cm$^2$/(Vs), respectively. In terms of charge disorder, at room temperature, epigraphene is in puddle regime for doping levels $|n|$<5x10$^{11}$ cm$^{-2}$ thus setting the maximum $R_H$ attainable in our epigraphene samples.

Fig. 2(c) shows the linearity of $V_H$ at 100 mT up to 6 mA bias current, measured for highly (n=1.6x10$^{12}$ cm$^{-2}$) and lowly (n=4.5x10$^{12}$ cm$^{-2}$) doped devices. We find that for all carrier densities the current-voltage (I-V) characteristic is linear within 5% error for $I_B$< 2.5 mA. The non-linearity is expected to be due to self-heating. For all subsequent measurements we limit the bias current to below 1.5mA to ensure a linear I-V behavior within 2% error.

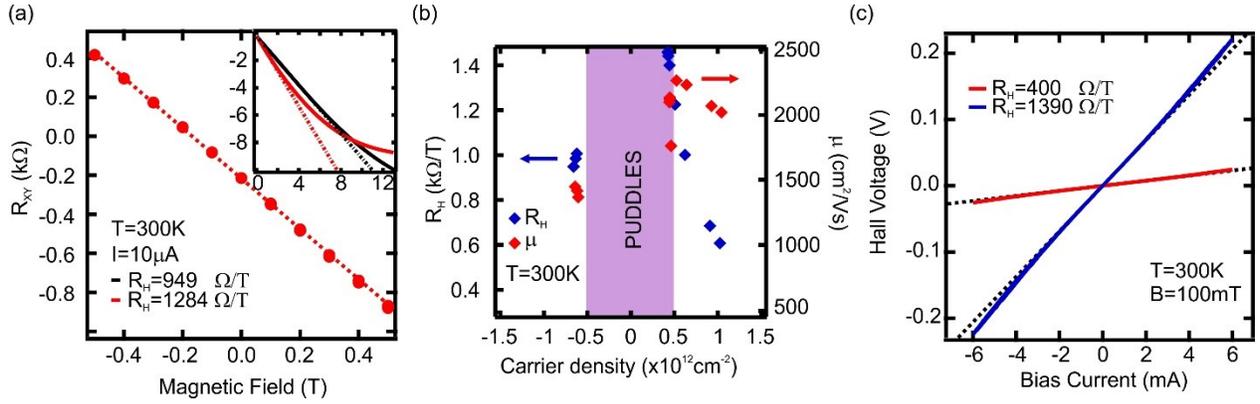

*FIG. 2. (a) Hall measurements showing linearity of $R_{XY}$ vs applied magnetic field. Inset shows behavior up to 13 T for different doping. The dotted lines are linear fits to low-field data $|B|<0.5$ T. (b) Carrier densities n and mobilities μ are extracted from low-field Hall measurements. (c) Linearity of Hall voltage measured at fixed field of 100 mT vs applied bias current. The dotted lines are linear fits to low-bias data $|I_B|<0.5$ mA. The offset in $V_H$ at zero field can be compensated by orthogonal VdP measurements and spinning current.[29] Typically observed offsets are on the order of 1 mV for a bias current $I_B=10$ μA (Supplementary S2).*

The measurements in magnetic fields are complemented with noise measurements to unveil the minimum detection limit $B_{MIN}$. Fig. 3(a) shows the low-bias ($I_B=10$ μA) voltage noise spectral density $V_N$ measured at the Hall voltage terminals for different doping levels. In the low bias regime, the corner frequency of 1/f noise is around ~30 Hz. As epigraphene approaches the Dirac point, the sheet resistance of the devices increases as $\rho_{XX} \propto 1/n$, and consequently the larger input and output resistance of the devices increases thermal noise. Dotted lines in Fig. 3 are the thermal voltage noise $V_{TH}$ calculated using measured input resistance. The agreement with experimental noise data points to the fact that, at low bias, thermal noise dominates in our sensors. Fig. 3(b) shows the increase of the 1/f noise contribution at larger bias currents, which nearly follows the Hooge's empirical relation with Hooge parameter $\alpha_H \approx 0.015$ (Fig 3(b) inset),[30] implying that the excess noise is mostly due to resistance fluctuations. In practical devices, the excess noise can be alleviated by using spinning Hall current measurement techniques.[29]

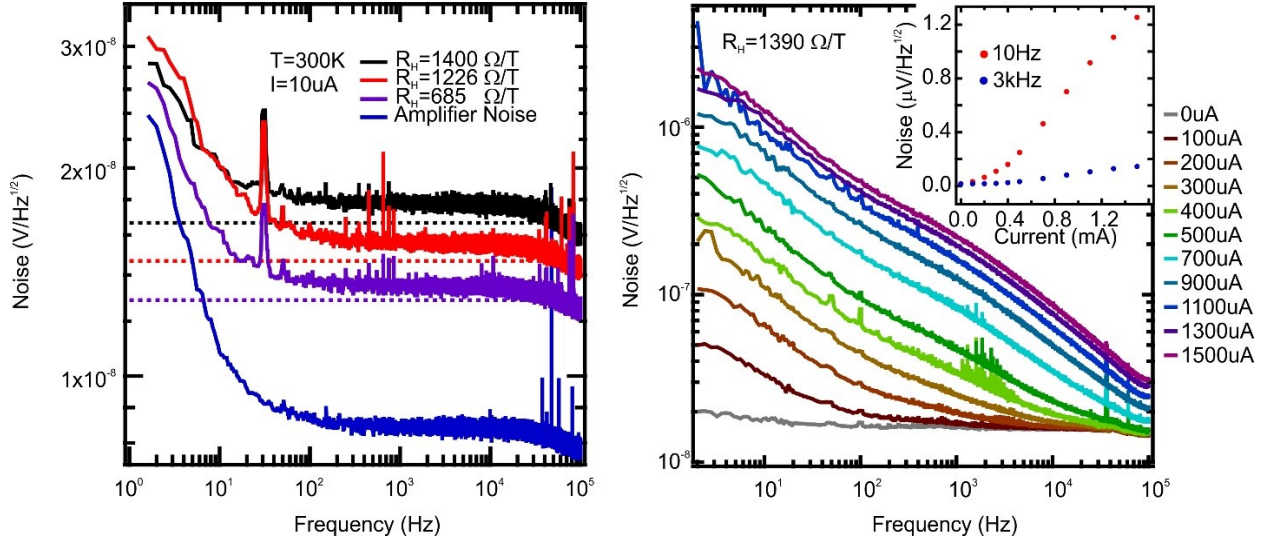

FIG. 3. (a) Noise performance for one Hall sensor measured at different doping levels. The dotted lines are calculated noise levels assuming pure thermal noise of a resistor. (b) Measured voltage noise spectral density vs bias current in another lowly doped device. Inset: The noise amplitude vs bias current at two different frequencies.

The measured sensitivities for epigraphene Hall sensors and their dependence on doping, collected across all measured devices, are summarized in Fig. 4(a). The highest $S_I$ is reached for low doping levels, close to the puddle regime ($n\sim 5\times 10^{11}$ cm$^{-2}$). The highest $S_V$ occurs slightly outside the puddle regime, at doping levels $n\sim 6\times 10^{11}$ cm$^{-2}$. We have performed full noise spectrum characterization (e.g. Fig. 3b) for four doping levels to obtain $B_{MIN} = V_N/(I_B R_H)$, which includes not only intrinsic noise of epigraphene (thermal and 1/f noise) but also amplifier noise. Fig. 4(b) shows $B_{MIN}$ as a function of $I_B$, measured at a frequency of 3 kHz for fair comparison to other graphene devices reported in literature. The best $B_{MIN}$=47 nT/√Hz is attained at lowest doping $n\sim 5\times 10^{11}$ cm$^{-2}$, for $I_B$=400 μA. At higher frequencies, where the 1/f noise contribution is lower, $B_{MIN}$ can be naturally lower with $B_{MIN}$=27 nT/√Hz, for $n\sim 5\times 10^{11}$ at 20 kHz (inset Fig. 4(b)).

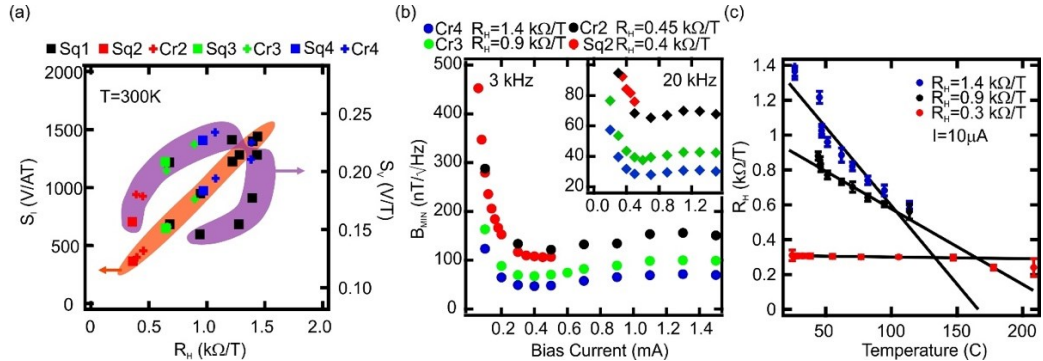

FIG. 4. (a) $S_I$ (orange region) and $S_V$ (purple region) versus $R_H$ compiled from 7 Hall sensors across 4 chips (Sq=square shaped, Cr=cross shaped). The sequence of data points span high to low doping (starting from the leftmost point). (b) $B_{MIN}$ versus bias current calculated directly from measured noise data for 3 kHz. Inset also shows data for 20 kHz. (c) Investigation of Finally, we describe the thermal stability of the molecular-doped Hall sensor through the temperature coefficient $\Delta_T$, defined as the percentage change of $R_H$ from its room temperature value per degree Celsius. Fig. 4(c) shows that samples doped close to neutrality ($R_H$=1400 Ω/T) are stable up to T=80 °C (Supplementary S3), with a temperature coefficient $\Delta_T$=-0.6%/°C. We achieve highest thermal stability with samples annealed for ~4 hours at T=160 °C, after which the $R_H$ reached a stable value of $R_H$~300 Ω/T due to partial desorption of dopants.[23] After this curing step at 160 °C, samples showed a fairly low $\Delta_T$=-0.03%/°C up to T=150 °C, while still displaying respectable performance at this temperature, with $S_V$~0.12 V/(VT), $S_I$~300 V/(AT), and $B_{MIN}$~100 nT/√Hz.

| Type | $S_I$ (V/(AT)) | $S_V$ (V/(VT)) | $B_{MIN}$ (nT/√Hz) | Freq (kHz) |
|---|---|---|---|---|
| InSb[29,31–33] | 140-700 | 1-7.2 | 1-60 | 0-50 |
| GaAs[29,31–33] | 30-3200 | 0.6-1 | 10-6000 | 0-50 |
| hBN-G[11] | 4100 | 2.6 | 50 | 3 |
| CVD[14] | 2093 | 0.35 | 100 | 3 |
| CVD[12] | 1200 | N/A | 300000 | 3 |
| CVD[13] | 97 | 0.03 | 400000 | 1 |
| Epi[34] | 1021 | 0.3 | 2000 | 3 |
| **Epi (this)** | **1080** | **0.23** | **60, 40** | **3, 20** |
| **Epi (this)** | **1442** | **0.21** | **47, 27** | **3, 20** |

*Table I. Figures of merit for room temperature Hall sensor performance, compared between both graphene-based Hall sensors and III-V commercially available sensors.*

Table 1 shows a comparison of our devices with other Hall sensors reported in literature. The maximum current-related sensitivity in doped epigraphene is found to be on the order of $S_I$~1.500 V/(AT) at room temperature. This value is limited by minimum $n$ attained in our sample ($|n|$<5x10$^{11}$ cm$^{-2}$), and is set by the disorder present in the as-grown material, combined with additional contributions from external doping and thermally excited carriers in the dopant layer and the SiC substrate. Decoupling epigraphene and substrate by hydrogen intercalation has led to high μ at cryogenic temperatures. However, at room temperature, the lowest $n$ reported for H-intercalated epigraphene are all above 1x10$^{12}$ cm$^{-2}$, with μ~1300-1700 cm$^2$/(Vs).[35] These mobilities are lower than the highest reported for epigraphene at room temperature (μ=5500 cm$^2$/(Vs))[22,36] and the ones achieved in this work (μ = 2300 cm$^2$/(Vs)). Above room temperature, interactions between epigraphene and the substrate via longitudinal-acoustic and remote interfacial phonon scattering further degrades mobility. The stable temperature range (T< 80 °C) for samples doped close to Dirac point is determined by our current choice of doping method.[23] A high thermal stability up to T=150°C is achieved after curing the samples at a temperature of 160°C for 4 hours. The resulting temperature coefficient $\Delta_T = -0.03\%/°C$ could then be understood as the intrinsic thermal drift of epigraphene, and not due to desorption of dopants. This implies that by using an alternate thermally-stable doping scheme, epigraphene could well outperform Hall elements based III-V at high temperatures. [29,31–33] Our work paves the way for development of epigraphene Hall sensors for real-world applications which require durable, controllable and sensitive devices produced in a scalable way.

We thank Alexander Tzalenchuk for insightful discussions. This work was jointly supported by the Swedish Foundation for Strategic Research (SSF) (nos. GMT14-0077, RMA15-0024), Chalmers Excellence Initiative Nano, and VINNOVA (nos. 2017-03604 and 2019-04426). This work was performed in part at Myfab Chalmers.

The authors declare that the main data supporting the findings of this study are available within the article and supplementary information. Additional data are available from the corresponding author upon request.

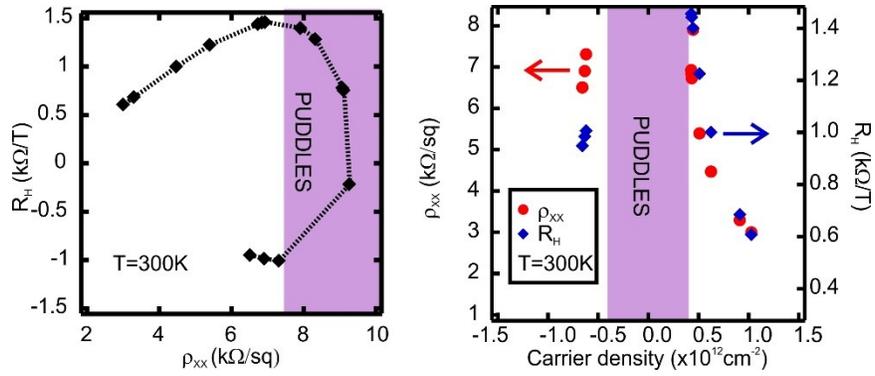

Fig. S1. Left: Relationship between Hall coefficient $R_H$ and sheet resistance $\rho_{XX}$. Sheet resistance is the highest in the puddle regime, while $R_H$ is the highest just outside the puddle regime. Right: $R_H$ and $\rho_{XX}$ vs carrier density n. $\rho_{XX}$ scales with carrier density $n$ in the same way as $R_H \sim 1/n$. Note that several $R_H$-values in the left plot have been omitted from the carrier density x-axis in right plot because the carrier density $n=1/(eR_H)$ is not well-defined in the puddle regime.

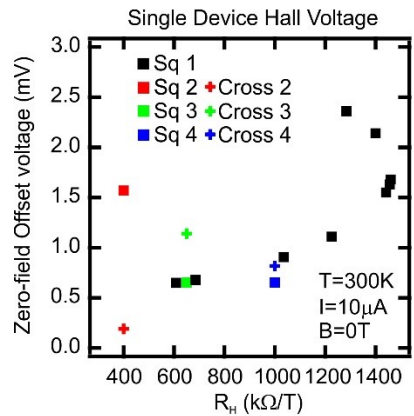

Fig. S2. Left: Uncompensated offset in Hall voltages at zero magnetic field measured for 7 different devices. In general, the offset voltages are on the order 1 mV, and tends to increase as samples are doped towards neutrality (puddle regime). Note that we estimate that the residual magnetization of the coil magnet is on the order of ~mT, further skewing the data to high offset values. In this limited dataset there is no observed correlation between device geometry and offset. The lowest offset voltage is achieved for cross geometry and high doping levels. Offset compensation can be achieved using orthogonal coupling of two or more Hall elements, in combination with Van der Pauw averaging, and can reduce the final offset to below 1 μV. Note that this requires very homogenously doped devices, which we do achieve when using molecular dopant F4TCNQ mixed with PMMA.

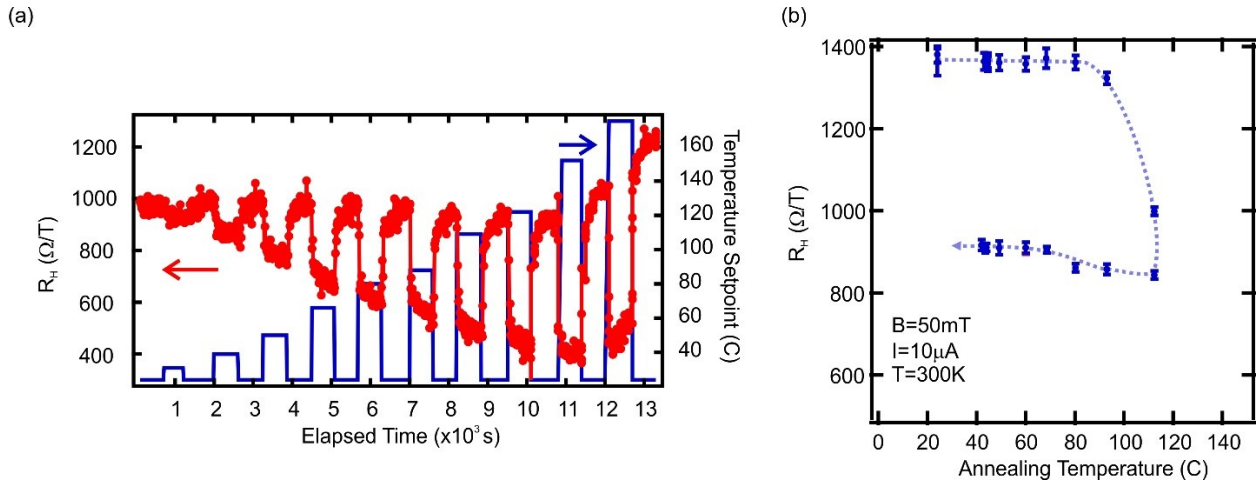

Fig. S3. (a) Example of investigation of thermal stability of a Hall sensor using in-situ heating sweeps. The red curve shows the Hall coefficient versus time and the blue curve shows the temperature set-point versus time. Sample behavior is monitored in real-time during heating and cooling. Fast sweeps (< 1 min) of magnetic field (±10 mT) are used to deduce $R_H$. The blue curve shows the temperature set-point versus time. (b) We study the thermal stability of the molecular-doped Hall sensor by measuring the room temperature performance of devices after repeated annealing. Initially, the devices is doped close to the Dirac point ($n<5\times10^{11}$ cm$^{-2}$) and is kept at elevated temperatures for 15 min, and left to cool down back to room temperature. Performance is assessed in real-time like in (a). This heating process is repeated, moving to successively higher temperatures. Only when the sample is annealed to above 80 ˚C, close to glass transition temperature of the polymer using for doping, does significant permanent change of room temperature *n* occur. There is a permanent increase in n-doping leading to a decrease in $R_H$. Subsequent heating above 80 ˚C induces further permanent change in doping toward even higher n-doping.